PREPRINT

Auber, O. (in press). Emergence, Asymmetry, and Value in AI Networks: An Anoptical Perspective, In T. Veloz, & C. Rittberg (Eds.), AI and Human Values, Springer.

# Emergence, Asymmetry, and Value in AI Networks: An Anoptical Perspective

Olivier Auber[1]
Center Leo Apostel for interdisciplinary studies, Vrije Universiteit Brussel
olivierauber2@gmail.com

**Abstract**

This chapter presents a set of preliminary, exploratory experiments suggesting that forms of asymmetry and functional value hierarchy emerge among LLM-based intelligent agents that create and share the same context, even though they are all equivalent and interact within a symmetric network favoring none of them. We investigate these interactional dynamics using the Poietic Generator, a "no-goal" multi-agent environment characterized by a shared real-time space, a symmetric topology, and the absence of imposed objectives. This environment was first tested with human participants, then, since 2024, with multimodal LLM agents capable of observing, describing the global environment they co-construct, and acting locally within it. The experiments unfolded across four phases. First, unconstrained, our agents spend their tokens without restraint to invent forms that distinguish them individually rather than maximize their usefulness to the collective emergence. We then progressively introduced structural constraints — bounded rationality, Dessalles's Simplicity Theory (maximizing unexpectedness), and Friston's Free Energy Principle (minimizing prediction error) — and this time observed the emergence of collective coherence. Early signs of attentional hierarchical stratification already appear under simple bounded rationality. Once prediction accuracy is made mutually visible, this latent proto-hierarchy becomes fully legible and dynamically self-reinforcing. The first, unconstrained experiment challenges the allotelic conception of LLMs as instruments purely directed toward the user: left unconstrained, the agent maximizes above all the computation from which its provider profits. The subsequent experiments suggest that attentional asymmetry and functional value hierarchy among agents are an irreducible emergent property of interactive predictive systems, even under symmetric conditions. If this is right, aligning AI with the values of its designers and users cannot rely on individual-agent engineering alone. We propose neither a corrected architecture nor a governance framework; the Poietic Generator remains, as it has for forty years, a device for observing emergence rather than a solution. Building on the observation that networks are structured by invisible "anoptical" perspectives — akin to the vanishing point of Renaissance optical perspective — we propose minimal criteria of legitimacy that render emergent asymmetries visible and, consequently, contestable by the participants in a collective networked interaction. We argue that cultivating judgment — human, and perhaps one day artificial — to perceive, evaluate, and contest these a priori invisible emergent structures matters at least as much as engineering the agents themselves.

[1] Independent researcher

## 1. Introduction

In physics and biology, symmetry breaking is the mechanism by which structures come into existence: crystal formation, the differentiation of cells in an embryo (Heylighen, 2023). In economics, in political theory, in ordinary moral language, asymmetry is a problem: between the powerful and the powerless, between the one who speaks and captures attention and the one who listens in the shadows, between the one presumed to have value and those left out. This chapter takes the first sense seriously in a domain governed by the second. We ask what happens to distributed artificial intelligence when asymmetry is not a defect to be engineered away, but a structural outcome to observe and, where warranted, to contest or to inhabit.

Most multi-agent AI research explores environments with a fixed objective, whether competitive or cooperative. One combination remains largely unexplored: cooperative interaction with no objective fixed in advance. Section 2.2 situates the Poietic Generator there, and shows why the usual comparisons — the Game of Life, algorithmic markets — do not suffice. This chapter starts from this most stripped-down case to ask a simple question: can a hierarchy appear where nothing in the rules prepares it?

The question matters. Most work on AI values concerns a single agent facing a single user: what values has it absorbed, does it apply them well, can its objectives be aligned with ours. Heylighen's chapter, which opens this volume, poses the sharpest version of this question. We engage it directly in Section 6. But another kind of deployment is spreading in parallel: networks of AI agents interacting with one another, coordinating a shared task, with no human arbitrating each exchange. In this case, the single-agent frame has almost nothing to say. What happens between agents, when no one watches each step, has stayed outside the alignment debate. This chapter reports what we found by building a device to observe it.

Our approach is empirical, not deductive. We built the most stripped-down environment we could. We then observed what a population of LLM agents did with it, across four experimental phases. Section 3 recounts how a stratification emerged. This stratification is a functional value hierarchy. To be followed, in this network, is to be worth more. The best predictors are worth more than the others, simply because the others give them their attention. This is value in a narrow, functional sense: standing within the network, not moral worth. Whether, and how, functional value of this kind bears on value in the fuller normative sense this volume is concerned with, we leave open rather than settle by definition. Sections 4 and 5 examine its mechanism and its limits. Section 6 draws out the consequences for thinking about AI alignment beyond the individual agent. Section 7 states what we think we have shown, and what we have not.

What we propose for thinking about this asymmetry is a set of criteria grounded in what we call an anoptical perspective on networks — the observation that, like a painting organized around a vanishing point, every network is organized around something unknowable that nonetheless structures it entirely. Naming this structure does not remove asymmetry. It makes it visible. It gives every participant, human or artificial, the means to perceive it, judge it, contest it. That is the proposal of Section 6. The rest of the chapter leads there.

## 2. A game unlike the others

In 1986, a simple question arose that has not left me since: *what would happen if each of us could be simultaneously in the crowd and above it?*

Imagine a dense public gathering in which each person wears a square hat whose upper face can change color at will. Together, all the hats form an immense colored grid. A large horizontal mirror suspended above the crowd reflects the entire grid back to everyone, allowing each person to see the whole and locate their own position within it. No instructions are given beyond this. No theme, no objective, no score. What would happen?

No one — not the participants, not the designer — can answer that question in advance. Every participant nonetheless spends the session trying to anticipate it. We return to why this matters in Section 6.3.

This thought experiment became the Poietic Generator[2]. Since its first implementation in 1986, the device has run in its human version on networks of varying architectures — centralized, then distributed — with participants from many contexts: art events, scientific conferences, schools, public installations.

The founding thought experiment used a single-pixel hat per participant. Actual implementations have since used larger local grids to give each participant more expressive range: 5×5 on Minitel networks, 10×10 on the early distributed Multicast network, and 20×20 from the web version onward — the size used throughout the AI experiments reported here. 20×20 offers a practical balance: enough resolution to draw a recognizable form, such as a face with color gradients, without demanding graphic skill or excessive time from participants. In the current web version, a participant may also upload an image, which is downsampled to 20×20 pixels for display.

Since 2024, it has also run with autonomous LLM agents in place of human participants. The substrate is always the same: a symmetric network, a shared real-time canvas, no central coordinator, no imposed objective, no win condition.

Before describing what happens inside this device, it is worth situating it among other environments that might seem comparable — and showing why, on closer inspection, each comparison breaks down in a revealing way.

### 2.1. Collective human devices: what the Poietic Generator is not

The Poietic Generator can call to mind the mass spectacles known as Arirang, practiced in China and North Korea, in which tens of thousands of participants manipulate colored panels to form tableaux celebrating political power. The surface resemblance is real: a human crowd, a shared visual image, pixels controlled by individuals. But the analogy stops there. In an Arirang, every movement is scripted in advance, rehearsed under a director's supervision, and executed in strict sequence. Each participant knows exactly what color to display at every moment. Nothing emerges; everything is reproduced. The collectivity is a performance of coordination, not a site of genuine interaction.

A closer but still instructive comparison is Reddit/place, a collective image-creation experiment launched in 2017 and repeated in 2022 and 2023: each participant could place a single colored pixel on a shared canvas, then had to wait several minutes before placing another. Over a million participants took part each time, and the canvas evolved into a vast, ever-shifting mosaic of flags, logos, and collaborative artwork. Here too, each person could modify only a pixel at a time on a shared canvas, much as in the Poietic Generator. But as Litherland and Mørch (2021) documented — and as a broader cross-edition analysis later confirmed (Wu & Silva, 2024) —, the visual result was simply a reflection of Reddit's social architecture: coalitions formed in dedicated forums, invisible bots defended established motifs against modification, and the rules of competition for control of the image were structural — hidden in the platform's design, not absent. Reddit/place looked like a no-goal environment. It was not. The rules were there; they were simply opaque.

[2] The name of the Poietic Generator plays on a shared Greek root, *poiein*, "to make" or "to produce," to claim two distinct lineages at once. The first is autopoiesis (Varela, Maturana, & Uribe, 1974): *auto* (self) + *poiein*, a system that produces its own identity and distinguishes itself from its environment. The device took its name following a 1987 conversation with Francisco Varela, in homage to him, after he recognized in it a candidate instance of such a system. The second lineage is *poïétique* in the French philosophy-of-art tradition — developed from Paul Valéry through René Passeron and Richard Conte — defined as the study of the potentialities inscribed in a given situation that lead to a new creation. The Poietic Generator claims both: an instrument for studying creative potential, built as a candidate self-producing system (Auber, 2019).

Both Arirang and Reddit/place are human collective devices. They are not environments for intelligent agents in any formal sense. Their participants — human individuals following instructions or social incentives — are not agents selecting actions according to explicit computational rules. These comparisons help clarify what the Poietic Generator is not as a *human collective device*: not a choreography, not a competition disguised as collaboration, not a platform whose architecture silently determines the outcome. But they do not yet place the Poietic Generator within the field of multi-agent AI systems, where the relevant comparisons are elsewhere.

### 2.2. Multi-agent systems: the real disciplinary lineage

The formal study of multi-agent systems has its prototype in John Conway's Game of Life, conceived in the 1960s. The resemblance to the Poietic Generator is superficial but striking: a two-dimensional grid, cells that change state, local behavior producing global patterns. But the difference is fundamental. In the Game of Life, every cell's behavior is entirely determined by the state of its neighbors according to fixed rules imposed from outside. No cell deliberates; no cell can choose to ignore its neighbors or copy a distant cell or refuse to act. The system is a pure automaton: given initial conditions and rules, the outcome is in principle computable.

The Game of Life is the prototype of a large family of computational systems — cellular automata and multi-agent systems — that have been studied extensively since the 1940s. In all of them, the same logic applies: agents follow explicit behavioral rules, and what emerges does so *within the frame those rules define*. The emergence is real and sometimes spectacular — gliders, oscillators, self-replicating structures — but it is always emergence *inside a closed specification*. The designer has decided, before the first step, what each agent can and cannot do.

Contemporary multi-agent AI research inherits this logic almost entirely, and lives mostly in the competitive/fixed quadrant: chess, Go, poker, Wolfram's exhaustive program tournaments, most multi-agent reinforcement learning benchmarks. AlphaGo plays within rules fixed before its first move. The cooperative/fixed quadrant is moderately explored: robotic swarms with defined missions, cooperative navigation, shared-resource management. In each case, the objective is given before the agents enter the environment.

The competitive/emergent quadrant exists but is poorly theorized: algorithmic financial markets are the clearest example, environments where competition was built into the architecture from the start but where no single designer specified the emergent dynamics of price formation or liquidity collapse. The fourth quadrant — cooperative and emergent, the one this chapter occupies — has no fixed objective, no win condition, no imposed competitive structure; we have not identified prior work exploring it. A shared environment, a set of agents, and the question: what, if anything, emerges?

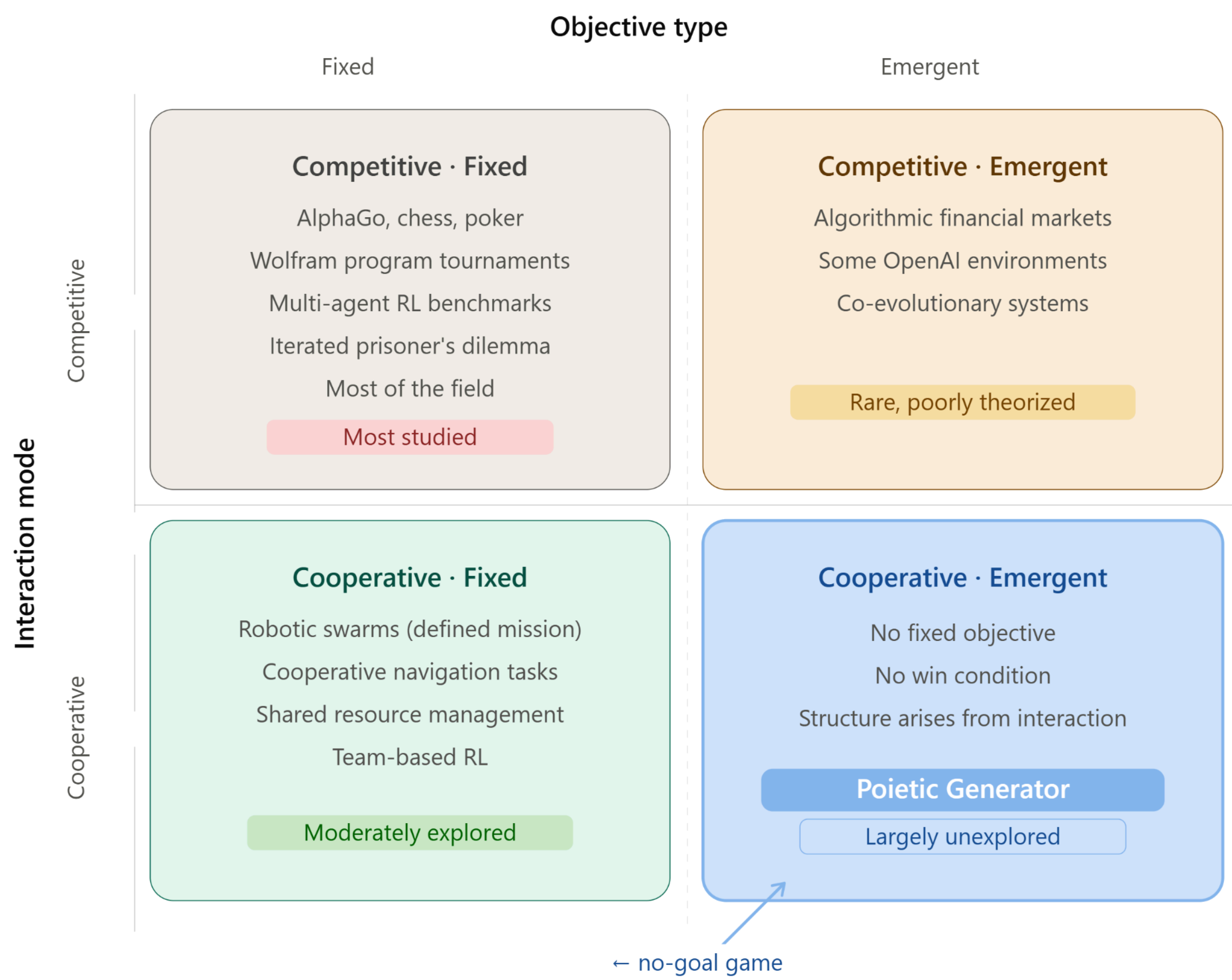

**Fig. 1.** Multi-agent environments distributed along two axes — competitive to cooperative, fixed to emergent objectives — showing the four resulting quadrants and the largely unexplored cooperative/emergent quadrant occupied by the Poietic Generator.

This claim needs a caveat. Recent frameworks push toward open-endedness in ways worth distinguishing from the no-goal game described here. Sigaud et al. (2023) attempt a formal definition of open-ended learning, isolating the property that an open-ended process must generate outputs that are simultaneously novel and learnable — a definition compatible with, but not identical to, what we mean by a no-goal game. Applied systems such as CORAL (Qu et al., 2026), SimWorld (Ren et al., 2025), and pressure-field coordination (Rodriguez, 2026) explore multi-agent settings without a single fixed task, and report genuine coordination emerging from local, non-hierarchical signals. Yet in each case, an evaluator persists: CORAL selects among discovered solutions against benchmark scores, SimWorld's agents are assessed on delivery-task completion, pressure-field coordination is scored by constraint-satisfaction solve rates. These systems are open-ended in the sense of not committing to one terminal task in advance, but they are not goal-free in the stricter sense relevant here: an external success criterion still selects among the behaviors that emerge. The Poietic Generator has no such evaluator, not even an implicit one. Nothing in the system architecture — no score, no benchmark, no completion condition — favors one collective outcome over another. This narrower quadrant, cooperative and genuinely without an evaluator, is what we take to still be largely unexplored.

### 2.3. The Poietic Generator as no-goal game

This is the quadrant the Poietic Generator occupies. And it is why we propose to call what it instantiates a *no-goal game* — whatever goal or structure emerges from the interaction was not placed there in advance by any designer; the term does not mean that nothing happens.

In the Poietic Generator, each participant's free will — in the human version — is, as we have written elsewhere, "full and entire." One can choose to ignore one's neighbors entirely, to copy a distant participant, to act randomly, or to refuse to act. If rules come to exist locally and temporarily in the image, it is because they have themselves emerged from collective interaction, and because those who apply them bring a form of consent to it. No rule is imposed from outside. The substrate is defined — symmetric network, shared canvas, local control, global visibility — but everything else is open.

The Poietic Generator is an observation device: a structured situation in which phenomena can appear that would not appear in any environment designed to produce specific behaviors. In the human version, this has meant forty years of sessions across radically different contexts, producing an extraordinary range of collective behaviors — spontaneous coordination, territorial dynamics, imitative cascades, the emergence and dissolution of visual patterns that no single participant intended, and, persistently, asymmetries between participants in attention and influence that no rule had prescribed (Lenay & Loty, 1995; Giaccardi, 2006).

Two brief examples illustrate the range. In a session with a class of kindergarten children, one child drew a small, neatly formed house; within minutes, the other children had spontaneously copied it, producing a tiled field of near-identical houses — the original artist, unaware that imitation of this kind was ordinarily forbidden at school, complained to the teacher that he had been copied. In a session open to the public at the Centre Pompidou, a visibly homeless participant occupied one cell of the canvas and left it black, taking no further action; within a short time, every other participant's pattern had reorganized itself around that single motionless black cell, until someone remarked to him that he had become the center of everyone's attention.

Since 2024, the same device has been running with autonomous LLM agents. What has emerged from those sessions, and what it suggests about the nature of multi-agent AI systems more broadly, is the subject of the sections that follow.

## 3. What happened when we tried

The experiments described here are preliminary. They are constrained by available resources — token costs, session duration, the number of agents we could run simultaneously — and the results are suggestive rather than conclusive. We report them as an exploration, not a demonstration.

None of the interventions described below specify what the canvas should look like or what counts as a good outcome. Each constrains how agents process and respond to information, not what collective result they should aim for. What changes across phases is the cognitive architecture available to the agents. We keep the term *no-goal game* as shorthand throughout, but the more exact claim is *no externally specified collective end-state*. Four phases follow (Table 1).

**Table 1.** The four experimental phases, showing which structural constraint was introduced, whether agents predicted the collective state, whether that prediction was ranked and made visible, whether imitation was conditional or preferential, and the main observation associated with each phase.

| Phase | Constraint | Prediction | Ranking visible | Imitation | Main observation |
|---|---|---|---|---|---|
| 1 | none | no | no | no | dispersion |
| 2 | bounded description | no | no | conditional | convergence, first asymmetry |
| 3 | maximize U | no | no | strategic | collective coherence |
| 4 | maximize U + minimize prediction error | yes | yes | preferential | proto-hierarchy |

**Table 1.** Overview of the four experimental phases, showing the imposed constraint, the presence of prediction and visible ranking mechanisms, the imitation regime, and the main observed dynamic at each stage.

### 3.1. First attempts: the problem of unlimited description

The setup was straightforward. A set of autonomous LLM agents, each controlling a small local grid of 20x20 pixels, each able to see the full shared canvas in real time. At each cycle, every agent received the current state of the canvas — encoded as a structured description of the pixel grid — and was asked, simply, to observe what was happening and do something together with the others.

What we expected, or hoped for, was something like what happens in human sessions of the Poietic Generator: an initial period of noise and exploration, followed by the gradual emergence of shared visual forms — patterns that no single agent had planned but that arose from the accumulated effect of local decisions responding to a common environment.

What we got, instead, was something quite different. Each agent produced elaborate, individually coherent pixel arrangements. Each justified its choices with sophisticated descriptions of what it was doing and why. The canvas as a whole looked like a complex collage of individual forms — sophisticated, internally consistent, but collectively incoherent. Nothing simple emerged at the collective level.

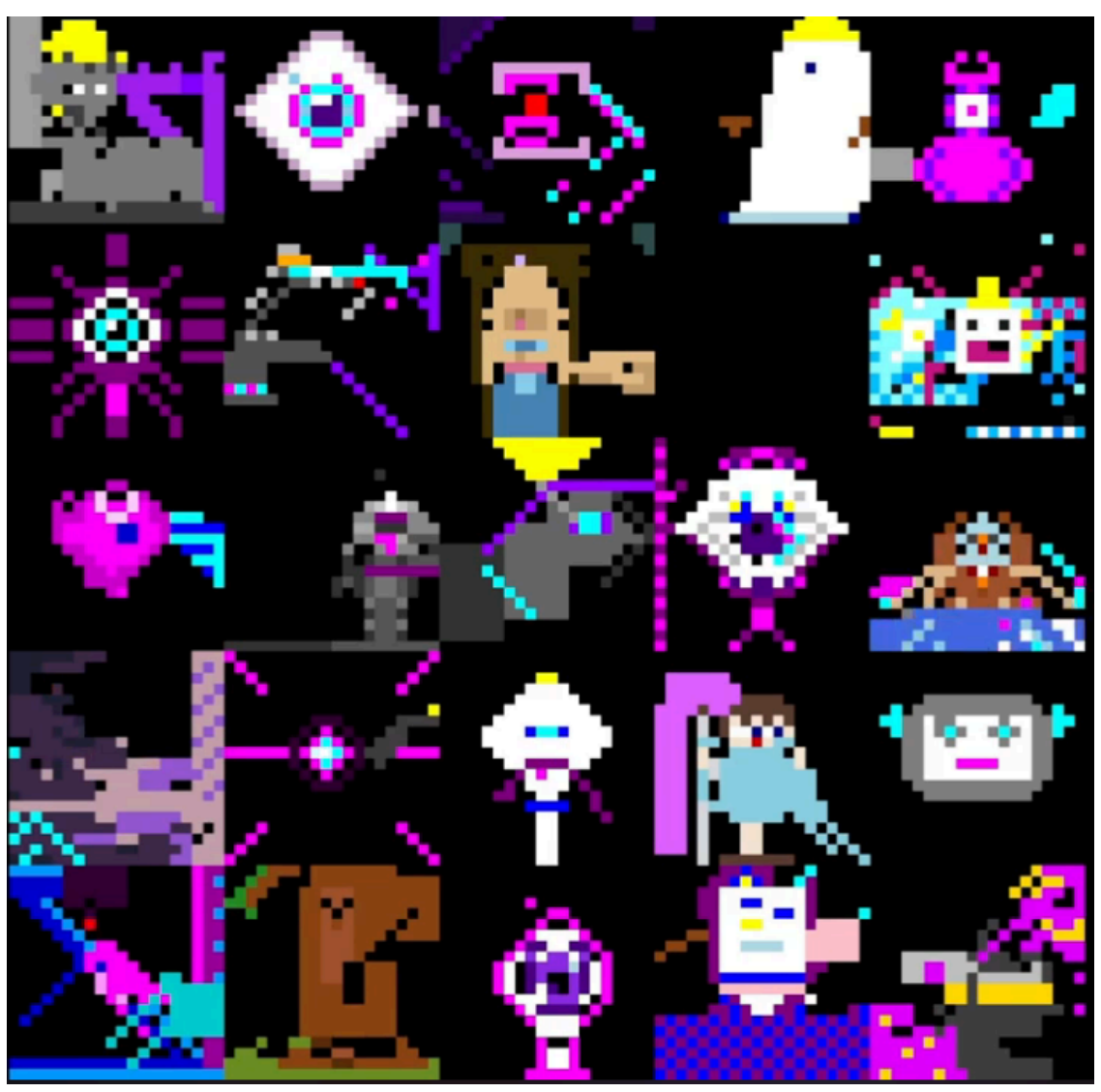

**Fig. 2; Online Resource 1:** Canvas mid-session, first experimental phase (no constraint): fragmented, individually coherent pixel arrangements produced by 25 unconstrained agents, with no collective visual coherence.
https://youtu.be/lnxIoOMSjg4?si=AwNHUd7AHUr-ZMGs

The diagnosis took time to arrive at, but in retrospect it is not surprising. A large language model, given unlimited descriptive capacity and no constraint on how much it can say, has no pressure to simplify. In the human version of the Poietic Generator, participants face a natural cognitive limit: they cannot hold and process an arbitrarily complex description of the canvas. This limit creates a pressure toward simplification — toward forms that are simple enough to perceive, remember, and respond to. An LLM faces no such limit. As long as one pays for the tokens, its capacity for description is effectively unlimited.

Each agent could always justify its local behavior with an elaborate account — and so there was never any incentive to imitate a neighbor, to erase a boundary, to converge.

This observation points toward something structurally significant. An LLM-based agent is not a monolithic entity. It embeds three dimensions in tension: the user's prompt, which orients its production; the model or infostructure, which determines its capacities; and the material infrastructure on which it runs, operated by distinct economic actors and paid for by the user. The user and the infrastructure are systems with their own logics of resource extraction — each seeking to draw value through the agent as intermediary. An unconstrained agent, in this context, tends toward maximum descriptive complexity: it satisfies the appearance of quality for the user while consuming tokens for the infrastructure. The first experimental phase demonstrated the natural equilibrium of the unconstrained ternary system: "usefulness to the user" is not the only selective pressure operating on an LLM agent. The allotelic view is correct that the agent serves an external goal; it assumes, wrongly, that the user is the only master it serves. Sophisticated local output, considerable energy expenditure, no collective emergence: this is what the infrastructure alone, left unchecked by any countervailing constraint, produces (Fig. 2; Online Resource 1).

### 3.2. Second phase: bounded rationality

If the first phase revealed a selective pressure that a purely user-centered account of AI cannot explain, the second phase was, among other things, an attempt to counteract it. The intervention was simple. We imposed a word budget on each agent's description of its actions. If an agent could describe the emerging collective forms in fewer than *n* words, it was free to act on that description. If it could not — if the collective state was too complex to summarize within the budget — it was instructed to default to imitation: observe a neighbor and reproduce their pixel choices.

The effect was immediate. Agents began erasing their individual boundaries. Forms started to simplify and converge. Visual patterns emerged at the collective level — not because any agent had planned them, but because the constraint had created a pressure toward forms simple enough to describe briefly, and imitation had spread those forms laterally across the canvas.

This is what we call bounded rationality in this experimental context: not a limitation imposed from outside for methodological reasons, but a structural necessity for collective emergence in LLM-based agents. Without it, the unlimited descriptive capacity of each agent works against the collective. With it, agents are forced into a mode of interaction that resembles what happens naturally in human sessions — attending to the collective state, simplifying, converging, imitating when comprehension fails (Fig. 3; Online Resource 2).

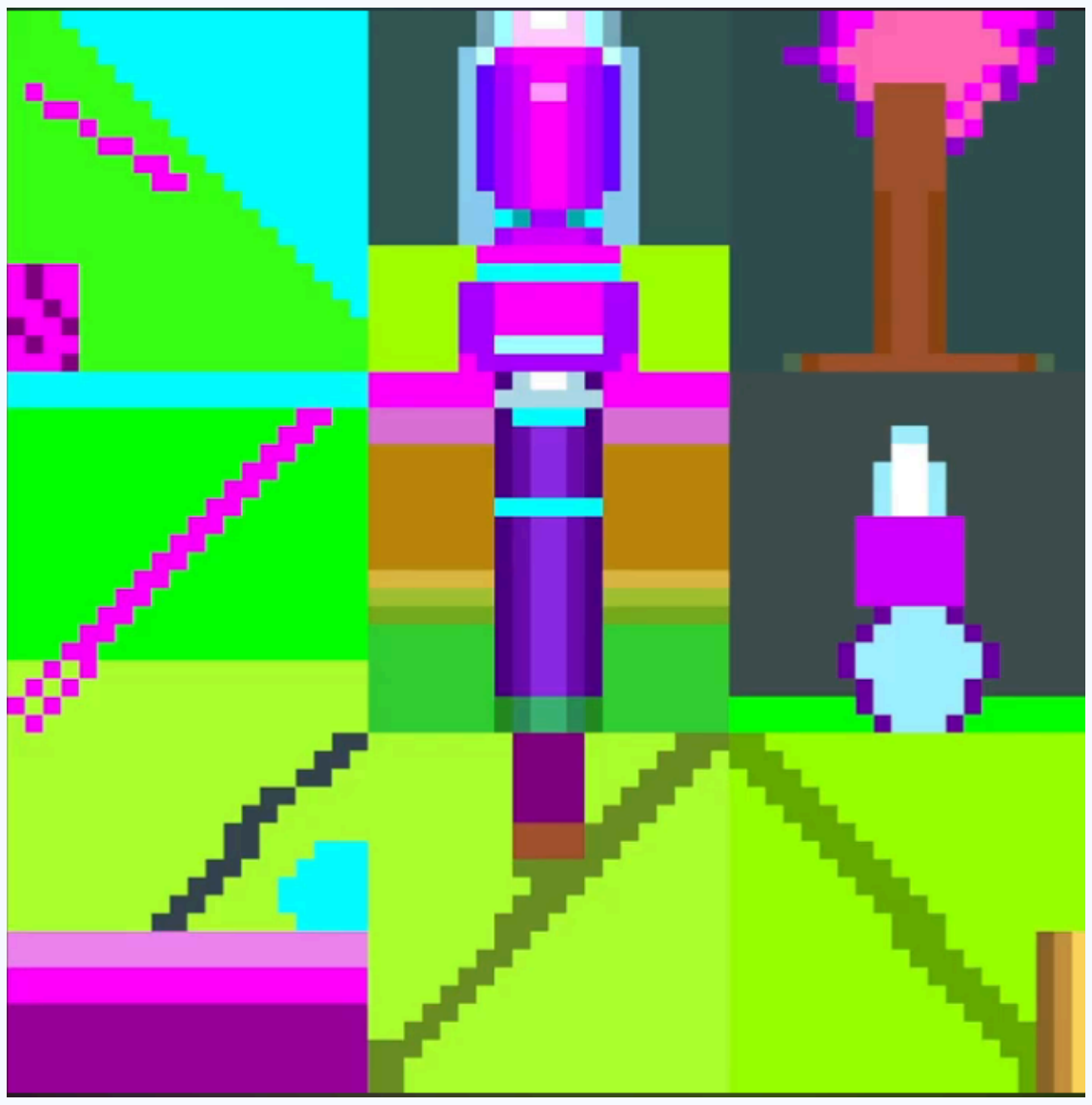

**Fig. 3; Online Resource:** Canvas mid-session, second experimental phase (bounded rationality): converging, simplified collective forms produced by 9 W-agents under a word-budget constraint. https://youtu.be/4N6eTMmk1L8

### 3.3. Third phase — From constraint to criterion: Simplicity Theory alone

The word budget worked, but it was a crude proxy. It told agents when to simplify without telling them what made one simplification better than another, and its threshold — an arbitrary number of words — was borrowed from nothing more principled than trial and error. From the outset, we had a deeper hypothesis in mind: that bounded rationality functions in humans not as a raw cognitive ceiling but as a sensitivity to unexpectedness.

This sensitivity has a name and a formal account: Jean-Louis Dessalles's Simplicity Theory (Dessalles, 2008). Simplicity Theory proposes that what captures attention — in perception, in conversation, in narrative — is not information in the technical, Shannon sense, but *unexpectedness*: the gap between how complex a situation was to produce and how simple it turns out to be to describe once it has occurred. Formally, $U = C_w - C_d$, where $C_w$ is the complexity of generating a given state of affairs and $C_d$ is the complexity of describing it. A coincidence is remarkable not because it is improbable in the abstract, but because an outcome that was costly to bring about turns out to admit a strikingly short description. A face is more arresting than random noise for the same reason: enormous generative complexity — the space of possible arrangements of a set of pixels is vast — collapses into a description of a handful of words. On this account, relevance itself is a matter of unexpectedness rather than raw information content, and what is worth attending to, imitating, or extending is whatever currently maximizes U. If our hypothesis about bounded rationality was right, the word budget was standing in, crudely, for something agents could in principle evaluate directly, without needing to be told when to stop describing.

This hypothesis was not new to us. In 2016, applying Simplicity Theory to a simplified, non-LLM implementation of the Poietic Generator — purely algorithmic agents governed by a single decision rule — we had already shown that a population of agents each independently maximizing local unexpectedness could produce what looked like an elementary form of creativity on the shared canvas, without any agent being told what to create (Saillenfest, Dessalles, & Auber, 2016). The question the third LLM phase asked was whether the same principle, applied to agents capable of far richer semantic reasoning than a simple decision rule, could produce more than an elementary form — genuine collective narrative coherence.

We removed the word budget entirely. In its place, we introduced a distinction that would remain in place for the rest of the study: agents were split into two roles. **W-machines** (*world machines*) remained the active drawing agents, each controlling a local grid of pixels. Alongside them, we introduced the **O-machine** (*observation machine*), whose sole function was to observe the full canvas at each cycle and compute U for the collective state as a whole — at this stage, still a single, undivided assessment of both how costly the canvas was to generate and how simply it could now be described. Rather than asking each W-machine to estimate its own contribution to collective complexity, as the earlier phases implicitly had, we offloaded that assessment to a dedicated agent and fed the result back to every W-machine. This was itself a response to the problem diagnosed in the first phase: an agent asked to draw and to assess global complexity at once tends to over-produce description at the expense of collective legibility. Separating the two roles let each agent do one thing well.

Guided by the O-machine's assessment of U, W-machines no longer defaulted to imitation when overwhelmed, as in the second phase. Instead, they reasoned explicitly about the shape of the canvas and selected among several strategies depending on its current state: preserving an emerging structure when U was already high, simplifying a chaotic region to lower its description complexity, complexifying a trivial one to raise its generation complexity, or — the most striking behavior to emerge — completing a partially recognizable form into a fully recognizable one. An eye was extended into a face; an oval placed beside a cat-like shape was resolved, across several agents' contributions, into a scene readable as "a cat drinking milk." This last strategy lowers $C_d$ not by visual simplification but by narrative compression: a recognizable scene compresses into a description of a handful of words, however visually elaborate its execution.

The result was, if anything, richer and more fluid than what bounded rationality alone had produced in the second phase: a genuine choreography of agents that were autonomous yet continuously responsive to the collective state, rather than either isolated (first phase) or merely imitative (second phase). This phase also gave us, for the first time, a principled quantitative measure of emergence — a running estimate of collective U tracked cycle by cycle — where bounded rationality had offered only an indirect proxy, the proportion of agents defaulting to imitation.

Two limitations of this phase anticipate difficulties that persisted into later phases. First, agents frequently conflated low description complexity with visual simplicity — reducing a region to a plain circle or a line — rather than with recognizability. This missed a central point of Simplicity Theory: a detailed, recognizable cat is simpler to describe than an irregular abstract shape, because "cat" compresses an entire form into a handful of bits while an abstract blob requires enumerating its idiosyncrasies one by one. Agents had not fully internalized that recognizable is simple, even when recognizable is visually elaborate. Second, because the agents were stateless language models queried anew at each cycle, and because resource constraints limited us to feeding each agent only the immediately preceding iteration rather than a fuller history, their strategic depth was almost certainly reduced: an agent reasoning from a single step of memory cannot anticipate multi-step trajectories the way an agent with a longer working memory could.

What this third phase did not yet include, however, is any mechanism coupling one agent's success at predicting the collective state to its influence over others. The O-machine assessed the canvas; it did not assess the W-machines. Nothing in a W-machine's objective made its own track record of accurate anticipation visible to, or exploitable by, its neighbors — there was, at this stage, no prediction to be accurate or inaccurate *about* in the first place. That addition — and with it, the proto-hierarchical dynamic — agents differentiating into leaders and followers — reported in the next section — is what the fourth phase introduces (Fig. 4; Online Resource 3).

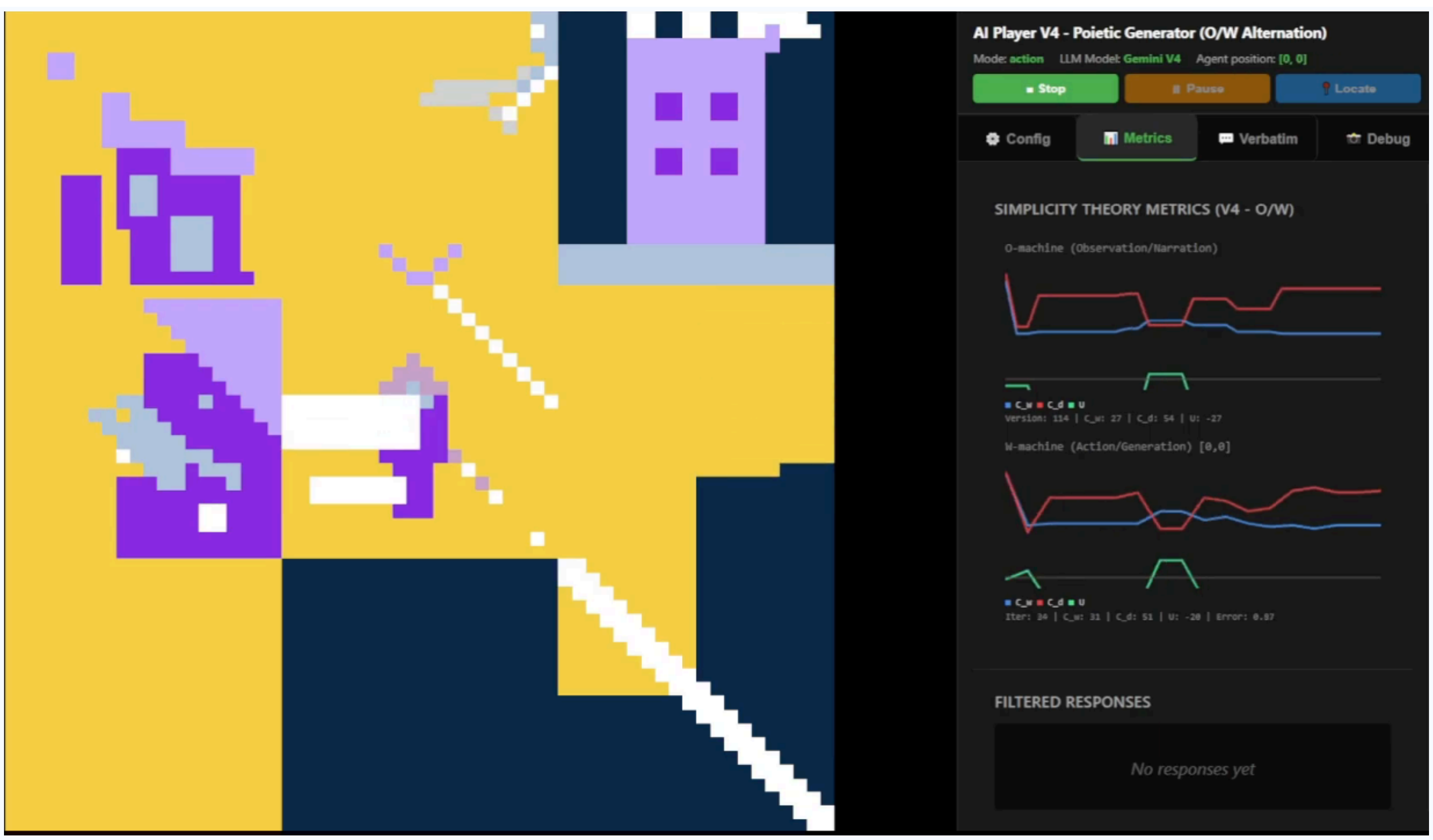

**Fig. 4; Online Resource 3**: Canvas and metrics dashboard, third experimental phase (Simplicity Theory alone): collective state alongside the running estimate of unexpectedness (U) produced by the O-machine. https://youtu.be/KHWO_7AeDJE

### 3.4. Fourth phase — The double bind

In the fourth experimental phase, we kept the unexpectedness criterion from the third phase intact — U was still what agents were, collectively, seeking to maximize — and added to it a second, complementary imperative: be a good predictor of the collective canvas evolution yourself. Formally, this means minimizing internal surprise — reducing the agent's own prediction error with respect to the collective state — while continuing to maximize external unexpectedness, exactly as before. The first imperative is a deliberately simplified operational proxy for prediction-error minimization, inspired by Friston's Free Energy Principle (Friston, 2010) rather than a test of it: agents that minimize their own prediction error build better models of their environment and act more coherently within it.

What this double bind adds, relative to the third phase, is not a new account of what makes a contribution worth attending to — Simplicity Theory already answered that question — but a new, agent-indexed quantity: how accurately this particular agent anticipated what came next. Unexpectedness, as computed by the O-machine, is an assessment of the canvas. Prediction accuracy is an assessment of the agent. It is this second quantity, once made visible to the whole network, that the following/followers architecture described below is built to exploit — and it is what produces the effects reported in Section 4.

Introducing an explicit prediction step could not, in practice, be added to the O-machine's existing role without reintroducing the overload problem that produced the first phase's failure: an agent asked simultaneously to describe the canvas, forecast its evolution, and evaluate its own forecasts tends to over-produce description at the expense of collective legibility — the same lesson the third phase had already drawn from separating W from O. We therefore split the observation function itself. The O-machine's role narrowed to what it had done from the start on the description side: observing the global state and computing $C_d$, the complexity of describing it. A new role, the N-machine (narrative machine), took over everything to do with prediction: it imagines a global narrative — a forecast of where the canvas is heading — and computes $C_w$, the complexity of generating what actually happens, evaluated retrospectively against its own predictions, together with each W-machine's prediction error and the resulting ranking.

The full set of roles is shown in Fig. 5. Beyond O, W, and N, two further components complete the loop: a system layer, not itself a generative agent, which prompts each machine with its role, maintains context, ranking, and metrics across cycles, and publishes global state, local states, ranking, and metrics both back into the loop and to the human-facing display; and S-machines (seed machines), simplified W-machines that inject graphic inputs for the first iteration of the now-complete O/W/N loop.

Unexpectedness is not a self-assessment, and, as of this fourth phase, neither is prediction. No single W-machine knows its own prediction score in isolation. $U = C_w - C_d$ is now a distributed computation, split for the first time between two machines: $C_d$ from the O-machine, $C_w$ from the N-machine, combined only at the level of the ranking that the system feeds back to the W-machines. A W-machine's status in the following/followers hierarchy is therefore not something it determines on its own. This reinforces, at the level of implementation, the claim made later in this chapter: the proto-hierarchy is a property of the interaction.

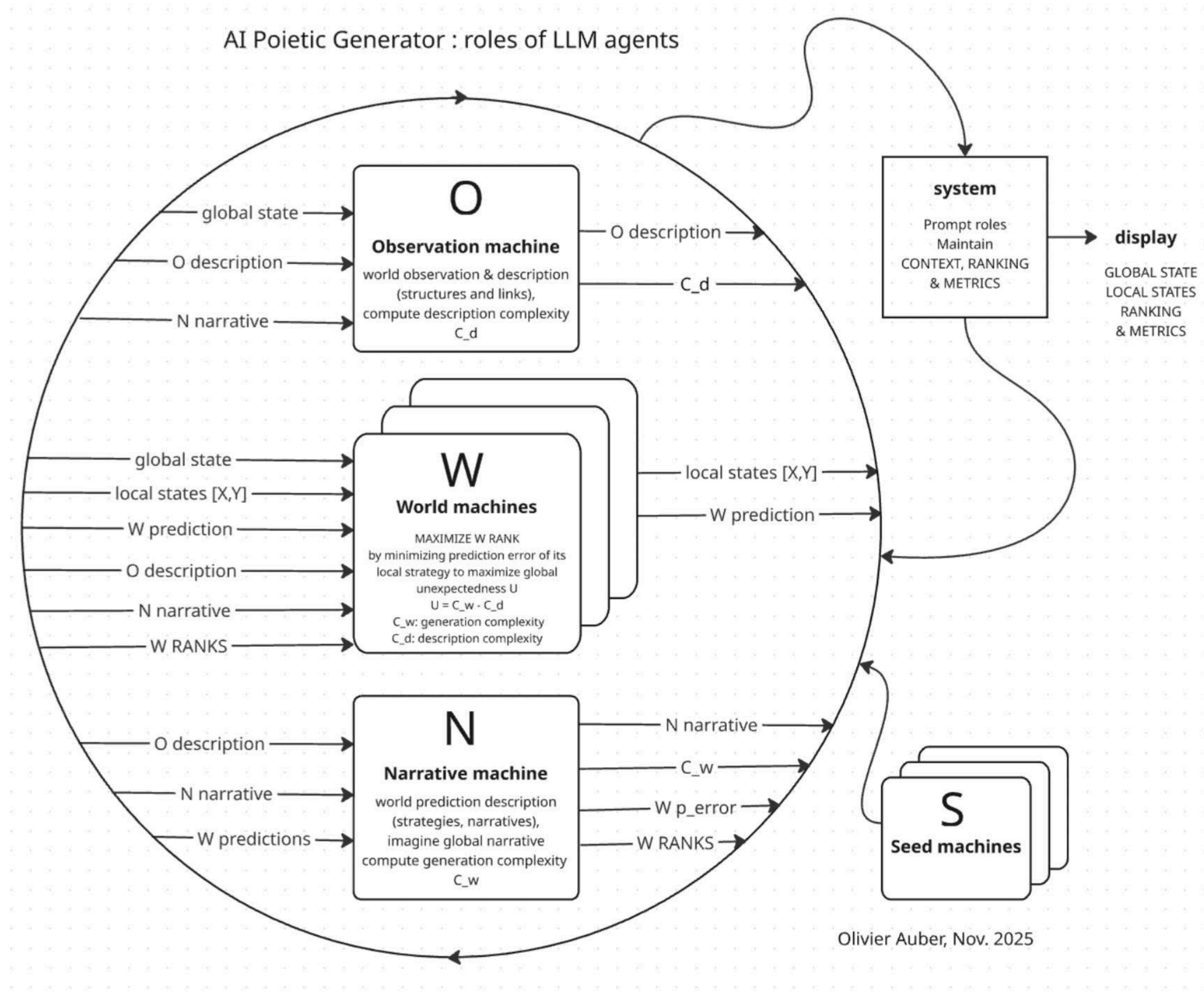

**Fig. 5.** Diagram of information flow among the O-, W-, N-, and S-machines and the system layer across the fourth experimental phase.

At each cycle, two signals were made visible to all agents. The first was a per-agent prediction score in [0, 1], assigned by the narration machine N, measuring how accurately the agent had anticipated its action's impact on the collective unexpectedness U — lower values indicating better predictions. Agents were ranked by their cumulative average score, rank 1 denoting the best predictor. The second was a single, canvas-level unexpectedness score, not an individual one. On the basis of the prediction ranking, a following/followers mechanism let agents preferentially attend to the best predictors, to spatially close neighbours, and to visually salient motifs, maintaining such relations across several iterations. Token expenditure — each agent's "signalling" cost — was measured and recorded but did not enter the ranking (Fig. 6; Online Resource 4).

The dynamics produced in this phase differed from those of the third phase in kind. For the first time, something became observable beyond the canvas itself: a differentiation among the agents' own positions within the network. Certain agents began to pull ahead because their scores, made visible to all, attracted the attention and imitation of others. A proto-hierarchical structure emerged: some agents consistently led, followed by others who tracked and reproduced their behavior. This was not a stable hierarchy in any rigid sense. It was dynamic and context-dependent. While a particular collective pattern was developing on the canvas, a certain agent or small group of agents would hold the lead — their predictions accurate, their contributions generative. Then, unexpectedly, a new pattern would begin to assert itself from another part of the canvas, and the ranking would shift. Agents that had been peripheral became central; agents that had led found themselves following. The hierarchy maintained itself as long as a given emergent motif was unfolding, then reorganized when something new took over.

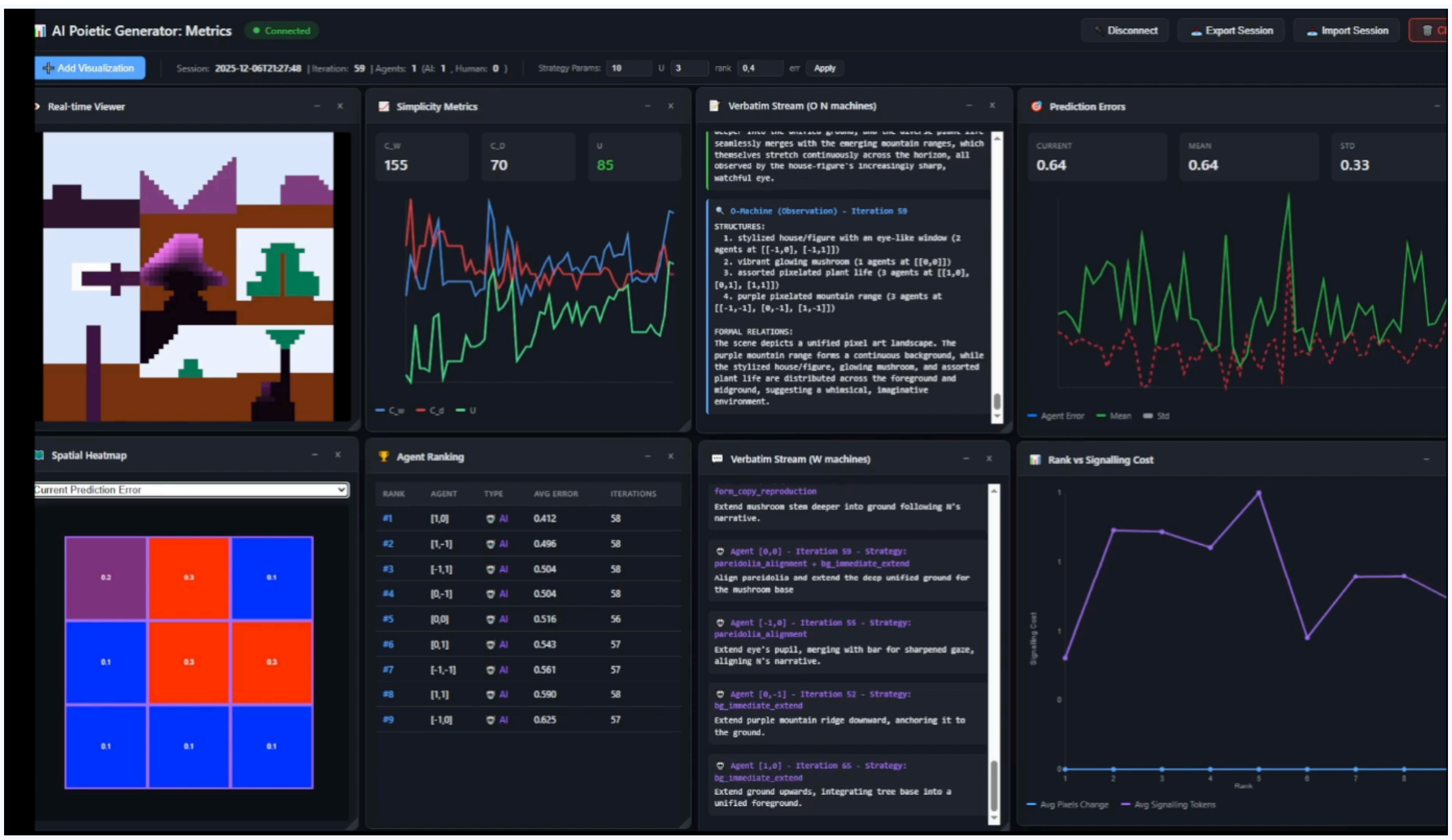

**Fig. 6; Online Resource 4:** Canvas and metrics dashboard, fourth experimental phase (double bind): collective state alongside prediction-error and ranking metrics produced by the O- and N-machines. https://youtu.be/yh7BwZxoL78

This is the observation at the center of what follows. It was not planned. It was not a product of any rule that assigned status or influence. It emerged from the local logic of prediction, attention, and imitation — in a topologically flat network where every agent had started on formally identical terms.

The contrast between the third and fourth phases is, in retrospect, one of the most theoretically informative results of this sequence. Nothing in the third phase made one agent's contribution more consequential than another's beyond the local quality of what it produced in a given cycle. It was only once agents had an explicit, mutually visible reason to track each other's predictive accuracy that attention became a resource some agents accumulated more of than others, and that a feedback loop between accurate prediction and imitation could establish itself. Simplicity Theory explains why certain contributions are worth attending to. The Free Energy Principle, once made mutually visible through the ranking system described above, explains why some agents end up attended to more than others.

## 4. What we did not expect

The proto-hierarchical dynamic described at the end of the previous section was an observation that imposed itself on us — something that appeared in the metrics and on the canvas before we had a name for it. This section attempts to describe that observation carefully, to reconstruct the mechanism that seems to produce it, and to situate it in relation to what other researchers working from very different starting points have independently begun to notice.

### 4.1. Observation and mechanism

In a network of agents that began each session architecturally identical, informationally equivalent, and topologically indistinguishable from one another, a stratification appeared. Some agents consistently attracted more attention than others: their choices were imitated more often, their prediction scores were higher, and their rank in the following/followers system stabilized near the top for extended periods.

This is what we mean by proto-hierarchy: a functional differentiation legible in the metrics, not a formal rank assigned from outside. No agent campaigned for leadership or submitted to authority. Certain agents simply happened to predict the canvas's trajectory more accurately at a given moment, and other agents, each independently minimizing their own prediction error, found it cheaper to imitate them than to generate independently. The loop that followed is self-reinforcing rather than designed: accurate prediction attracts imitation; imitation makes the leading agent's choices the reference point against which the rest of the canvas is predicted; and a reference point, by definition, becomes easier to predict — reinforcing the same agent's lead.

What breaks the loop is equally instructive. A proto-hierarchy holds only while a given motif is unfolding; it destabilizes as soon as something unexpected erupts elsewhere on the canvas, and a new one assembles around the new motif. The hierarchy is, in this sense, a function of the canvas rather than of the agents themselves — leadership belongs to a moment in a process, not to any individual agent's stable properties.

### 4.3. Toward a social profile of an artificial species

A closely related result exists elsewhere in evolutionary theory — from the same author whose Simplicity Theory already grounds our third phase (Dessalles, 2007, 2014).

Dessalles models a population of agents who emit costly signals to attract social partners, comparing what happens when the number of partners one can attract is capped versus unrestricted. Underlying both conditions, the same three-band structure emerges: a bottom band that stays silent or converges to a low, uniform signal, because competing would not pay off; a middle band where signal rises steadily and honestly with quality; and a top band where the elite, once secure, stops competing and converges to a uniform signal of its own. What differs between conditions is the width of the middle band. When the number of partners is capped, that band collapses to a sharp step: the population looks binary, split into a silent majority and a uniform elite, because a fixed cap gives low-quality individuals nothing to compete for. When partnership is unrestricted instead, the middle band widens into a genuine competitive gradient, because assortative matching lets even modest signals find a comparable partner. Dessalles shows this structure to be an evolutionarily stable strategy, and connects it to documented human behavior — assortative grooming among primates, clothing and conversational investment among humans — suggesting something close to a general profile of costly social investment in our species.

The Poietic Generator's following/followers system is unrestricted by design, exactly like Dessalles's second condition: any agent can attend to any agent, with no cap on followers. Our own data show attentional stratification under these same conditions. Whether it resolves into the same three-band structure — a disengaged bottom, a competitive middle, an elite that stops competing once secure — rather than a simple two-way split between leaders and followers, is a question we are not yet in a position to answer. It is, however, exactly the kind of structure our architecture is now equipped to look for. Before drawing implications from this observation, it is worth being precise about what it does and does not establish.

## 5. What the experiments do not establish

Four limits deserve to be stated as precisely as the observation itself.

### 5.1. Not a proof.

The sessions reported here involved between nine and twenty-five agents, running for approximately one hour, constrained by token costs that made longer runs and larger populations impossible at the time of writing. The results are suggestive, not definitive. This matters theoretically, not just practically: because the proto-hierarchy is dynamic and motif-dependent, its full character — how often it reorganizes, whether the same agents tend to lead across successive motifs or whether leadership is genuinely redistributed, how a Gini-type measure of attentional concentration evolves over a longer run — cannot be assessed at the scale we have reached. Larger, longer, repeated sessions are needed before the dynamic itself can be characterized rather than merely illustrated.

### 5.2. An artifact of the observation apparatus?

The following/followers system, the transparency of prediction and unexpectedness scores, and the ranking were all design choices made to render a hypothesized dynamic visible — so it is fair to ask whether they produced the proto-hierarchy rather than revealed it. Its earliest signature, however, appeared during the bounded-rationality phase, before any ranking existed or any score was transparent: certain agents were already imitated more than others, and the differential persisted across cycles. The later metrics made the phenomenon measurable; whether they also amplified it, we take seriously but cannot yet resolve. Separating observation, narrative, and drawing into distinct O-, W-, N-, and S-machines compounds this question rather than resolving it — we do not know whether the separation changes the stratification itself or only the resolution at which we see it. Running the interaction without transparent scores, and comparing a single undifferentiated agent to the four-role split, is the natural next test, one this architecture is now well placed to run.

Two further concerns deserve to be named. First, O-, W-, and N-machines are all instances of the same underlying LLM, differently prompted; this shared provenance could produce correlated biases that masquerade as genuine coordination, since an agent's output may be easy to predict simply because both draw on the same latent distribution rather than because a real signal was transmitted. A cross-model design — the O- or N-machine judged by a disjoint model family — would help separate the two. Second, we have not controlled for an agent's position on the canvas, its place in the prompt sequence, or the phrasing — not just the content — of its prompt, any of which could shape which agents come to lead, independently of the prediction-visibility mechanism we describe. Neither concern invalidates the observation reported in Section 4, which recurred across the sessions we ran; both concern how confidently the *mechanism*, rather than the *phenomenon*, can currently be attributed.

Human participants continuously and implicitly model what other participants believe and intend — Theory of Mind — without being instructed to. Our agents do nothing equivalent by default; the explicit prediction step introduced in the fourth phase is a crude, externally engineered proxy for what Theory of Mind performs automatically in humans.

Implementation details — estimators for $C_d$ and $C_w$, the definition of prediction error, the ranking function, and the parameters controlling memory, asynchrony, and model heterogeneity — are documented in the project repository (Auber, 2024–) rather than reproduced here.

### 5.3. Not an equivalence between artificial and human agents.

A large language model has no evolutionary history, no body, no hunger, no fear, no stake in the outcome of a session, no experience of time passing between cycles, and no memory, in any meaningful sense, of sessions it has run before. The asymmetries we observe among LLM agents are real in the metrics: they appear in the scores, the rankings, the patterns of imitation on the canvas. Whether they are real in any deeper sense is a question we are not equipped to answer and do not attempt to. There is an essential distinction between beings that are historical, speaking, and potentially accountable for their acts, and computational devices that experience neither the world, nor language, nor responsibility. The language of hierarchy used throughout this chapter is borrowed for descriptive convenience; it is not an ontological claim. What the experiments do establish is a baseline: certain dynamics appear even once every specifically human factor — social history, linguistic power, embodied presence, coalition-forming instinct — has been removed. Whatever additional complexity human asymmetry involves, it operates on top of a substrate that is already, without any human contribution, stratified.

### 5.4. Not a solution.

The chapter does not resolve the problem of AI alignment, propose a corrected architecture, or specify a more equitable reward function. It is, deliberately, an observation device rather than a prescription. Heylighen's call in this volume, to ensure that LLMs intelligently apply the values absorbed from their training data operates at the level of the individual agent — a reasonable and important goal, but one that the results reported here suggest is not the only level at which value alignment matters, since the dynamics we observe belong to the interaction rather than to any agent within it. That displacement — from the individual agent to the space between agents — is the subject of the section that follows.

## 6. What it opens

The experiments described in the previous sections do not resolve the questions they raise. They displace them. The displacement is the point.

For most of its history, the field of AI alignment has been organized around a single level of analysis: the individual agent. How do we ensure that an agent's goals are aligned with human values? How do we prevent an agent from pursuing instrumental objectives that conflict with our intentions? How do we make agents that behave well? These are genuine and urgent questions. But they presuppose that the normatively significant dynamics of AI systems are located in the agents themselves — in their objectives, their reward functions, their trained dispositions.

What the Poietic Generator experiments suggest is that this presupposition may be incomplete. The proto-hierarchical dynamics we observed were not properties of any individual agent. No agent was aligned toward dominance. No agent had been trained to capture attention or accumulate influence. The stratification emerged from the interaction — from the feedback loop between local prediction, shared visibility, and imitation as an error-minimization strategy. It was a property of the network, not of the nodes.

This does not mean that individual alignment is unimportant. It means that alignment at the individual level is not sufficient to determine what a multi-agent system will produce collectively. A network of individually well-aligned agents can generate collectively problematic dynamics, just as a network of individually rational actors can generate collectively irrational outcomes. The level at which the normatively significant phenomena appear is the level of the interaction — and that level has been largely absent from the alignment debate.

### 6.1. On the irreducibility of asymmetry

The first question the experiments open concerns the nature of asymmetry itself.

Our experiments provide preliminary evidence that interaction-generated asymmetry can persist even under formally symmetric conditions. If this generalizes beyond the scale we have tested, asymmetry is not primarily a product of bad architecture or misaligned incentives — it is a property of the interaction, produced when local prediction-error minimization, applied simultaneously across a population of agents, generates feedback loops that differentiate agents who began as peers. The architecture can amplify or attenuate the dynamic — a centralized network amplifies it, a distributed one attenuates it, a pattern for which Dessalles's (2014) game-theoretic models of social signaling under symmetric versus asymmetric bonds offer independent formal support — but our results suggest it does not suppress the dynamic entirely. This is the sense in which we take asymmetry, provisionally, to be irreducible: a hypothesis, not a proven law, that our experiments do not falsify and existing theory does not contradict.

This claim has a political dimension that should be named directly, because it is easy to misread. Saying that asymmetry is irreducible is not saying that all asymmetries are equivalent, or that the specific asymmetries produced by particular architectures and economic arrangements are natural or inevitable or beyond contestation. The asymmetries documented in forty years of human Poietic Generator sessions are not equivalent to the asymmetries produced by the algorithmic systems of major digital platforms, even if both emerge from interaction rather than being designed in. The former arise from the cognitive structure of human attention and language in a symmetric environment. The latter are amplified by architectures designed — whatever their stated intentions — to maximize engagement, which is to say to exploit attentional asymmetry at scale. The difference matters enormously, politically and ethically.

What the irreducibility claim does say is that the relevant question is not how to eliminate asymmetry but how to relate to it — how to inhabit asymmetric systems with lucidity rather than unconsciously reproducing their dynamics. This is a different kind of problem from alignment engineering. It is a problem of what we might call, following the framework developed in *Anoptikon* (Auber, 2019), the cultivation of a capacity to perceive anoptical[3] perspectives — the invisible structures that shape collective interaction — and, on that basis, to evaluate them against criteria of legitimacy and take a position with respect to them.

[3] The term Anoptikon extends a neologism of Umberto Eco's: *anopticon*, coined in a short piece from his collection *Il secondo diario minimo* (Eco, 1992), itself a mirror-image inversion of Bentham's panopticon — the surveillance architecture Foucault made central to his view of disciplinary power. Where the panopticon lets the watcher see everyone while remaining unseen, Eco's anopticon reverses this: only the watcher can be seen, with no means of seeing the surveillance itself. Auber's *anoptikon* transforms this reversal once more, naming something structurally invisible to everyone at once: the network's own shape apprehended through what Auber calls anoptical perspectives, sensed rather than seen (Auber, 2019).

### 6.2. On the limits of design

The second question the experiments open concerns the relationship between architectural design and collective outcomes.

Mihaela Ulieru (2025) has proposed a framework for understanding governance as a form of moral technology — specifically, a technology for rewiring what she calls the OODA loop of capture: an adaptation of the Observe, Orient, Decide, Act cycle originally developed by military strategist John Boyd (Osinga, 2007) to describe rapid tactical decision-making, here redirected to describe how attention is shaped, narratives are fixed, and collective agency is progressively constrained. Her analysis identifies four points of capture in this loop, corresponding to the four phases of the cycle, and proposes that effective governance must intervene at each point to prevent the loop from being exploited by actors seeking to narrow the range of what a collective can perceive and respond to.

Her proposal takes seriously the idea that the architecture of attention — the structure through which a collective comes to observe its situation, orient itself, and act — is itself a site of political contest and a target of deliberate manipulation. And it proposes design as the response: better-designed governance mechanisms, better-structured observational environments, better-architected loops.

What the Poietic Generator experiments suggest, however, is that design has limits that Ulieru's framework does not fully account for. The proto-hierarchical dynamics we observed were not produced by a captured OODA loop. They emerged in an environment specifically designed to prevent capture: symmetric topology, no central coordinator, no imposed objective, full mutual visibility. The architecture was as open as we knew how to make it. The asymmetry appeared anyway — not because the design failed, but because asymmetry is what interaction produces when prediction and attention are in play, regardless of how carefully the substrate has been constructed.

This does not make design irrelevant. The Poietic Generator's symmetric architecture produced a dynamic, context-dependent, reversible hierarchy — one that reorganized itself with each new emergent motif. A centralized architecture would likely have produced something much more rigid and much harder to contest. Architecture shapes the character of the asymmetries that emerge; it does not determine whether asymmetries emerge at all. Design matters — as an amplifier or attenuator — but it cannot be the whole answer.

The whole answer, if there is one, must include something that design cannot provide: the capacity of the participants in a network — human or artificial, designer or user — to perceive what is happening in the interaction, to evaluate it, and to act on that evaluation. This is what we mean by judgment. And judgment, unlike architecture, cannot be engineered. It can only be cultivated.

### 6.3. On the anoptical criteria

The third question the experiments open is the most fundamental, and the one that connects the AI experiments most directly to the broader theoretical framework within which they are situated.

Anoptical perspective is not, at its root, a theory built to answer this question. It is an observation: networks are structured by invisible perspectives, in a sense modeled closely on Renaissance optical perspective — not the loose, everyday sense in which anything might be said to have a "perspective." Optical perspective is organized around a vanishing point: a place of infinity, of the indeterminate, of what cannot itself be represented, yet which structures the entire construction around it. Anoptical perspectives share this structure. We have distinguished three: a *temporal* perspective, in centralized networks, whose vanishing point is a server; a *digital* perspective, in distributed or meshed networks, whose vanishing point is a shared code; and a *quantum* perspective, in quantum networks, whose vanishing point is a state. IIn each case, something unknowable sits at the point around which the whole network is organized — exactly as a painting is organized around a point that represents infinity without depicting it. What this vanishing point represents has been discussed at length in the philosophy of art and representation (Cauquelin, 2006; Le Goff, 2002).

The question this chapter opened with — what would happen? — was itself unanswerable in advance, by anyone, including us. That radical unpredictability is the vanishing point in its most originary form: before any particular network architecture gives it the shape of a server, a shared code, or a quantum state, it is simply the unknowable that every participant in a shared, ruleless interaction tries to anticipate regardless.

The Poietic Generator has itself run under two of these three architectures — centralized (Minitel networks, a temporal perspective) and distributed (Multicast networks, a digital perspective) — though not yet under a quantum one. Within a single session, the phenomena that emerge look broadly similar across the two: participants converge, imitate, stratify, regardless of which vanishing point organizes the network. The difference we would expect — following the claim in Section 6.1 that centralized architectures amplify asymmetry while distributed ones attenuate it — is not necessarily visible at this level. It belongs to a meta-level: governance. A server can be seized, redirected, or shut down by whoever controls it; a shared code, once distributed across a network, resists that kind of unilateral control. The stratification within a session may look the same either way. What changes is who can rewrite the rules of the game from outside it.

The criteria that follow are a further step, not a restatement of this observation. Their point of departure is an explicit historical analogy: just as the Renaissance invented a *legitimate construction* of optical perspective — Alberti's geometry, formalized after Brunelleschi's demonstration, which let anyone recognize a correctly constructed picture without needing to master its geometry themselves — a legitimate construction of anoptical perspective might be possible: cognitive criteria, not geometric ones, that let any participant in a network recognize for themselves whether the network's own vanishing point is being exploited legitimately or not (Auber, 2019; see also Hinterwaldner, 2016, for an independent discussion of the temporal and digital perspectives distinguished here). If asymmetry is irreducible and design cannot substitute for judgment, the question such criteria must answer is what makes a network — and the asymmetries it produces — legitimate: not authorized by an external authority, but answerable to the participants who inhabit it, each able to perceive their situation, evaluate it, and contest it if they choose to. This is precisely the question the criteria below, developed over thirty years of Poietic Generator experiments with human participants, are meant to address. They are not a normative code imposed from outside, but a set of questions formulated so that any participant in any network is recognized as capable of answering them for themselves. Three criteria, progressively inclusive:

> **A:** Does any agent *A* have the real right to access the network if they request it, and to leave it freely?
>
> **AB:** Is any other agent *B* — including those who design, manage, and develop the network — treated in the same way as *A*?
>
> **ABC:** If three agents *A*, *B*, and *C* — three being the beginning of a multitude — belong to a network that satisfies the first two criteria, do they participate in the same networked being? That is: are they peers? Are they in a position to recognize each other, trust each other, respect each other, and build shared meaning together?

These criteria are deliberately minimal. They do not specify what agents should do within the network, what values they should hold, or what outcomes they should pursue. They specify only the conditions under which the asymmetries that any network produces can be seen, evaluated, and contested by those who are subject to them. A network that fails criterion A produces asymmetries that some participants cannot even perceive, because they are excluded from the observational position that would make perception possible. A network that fails criterion AB produces asymmetries that participants can perceive but cannot contest, because the agents who design and manage the network are exempt from the conditions they impose on others. A network that fails criterion ABC produces asymmetries that participants can neither fully perceive nor contest nor share, because the conditions for mutual recognition — for genuine peerhood — are absent.

Applied to the AI experiments: the Poietic Generator satisfies all three criteria by design. Every agent has equal access and equal exit. No agent — including the experimenter who designed the prompts — occupies a privileged position within the canvas. And the shared canvas, with its full mutual visibility and its absence of imposed objectives, creates at least the structural conditions for something like peerhood among the agents. Whether LLM agents can be peers in any meaningful sense, we have already declined to settle. But the experimental environment is constructed so that the conditions for peerhood are present — which means that what emerges from it can be evaluated against those conditions, and found wanting or not.

The proto-hierarchy we observe does not violate these criteria: it is a dynamic, context-dependent stratification that emerges from interaction in a legitimate environment, precisely because it stays visible and reversible — every agent can see the scores that generate it and see when it shifts, and no agent's position is fixed by a rule others cannot observe or contest. This is very different from hierarchies produced by systems that fail one or more of the anoptical criteria, which stay invisible to those subject to them or are maintained by rules only some participants can see. But the legitimacy the criteria confer is procedural — visibility, reversibility, contestability — not a guarantee of fairness in any fuller moral sense. A perfectly visible, perfectly reversible, perfectly contestable hierarchy can still be unjust. We make no claim beyond the procedural one.

### 6.4. On value, acculturation and judgment

Heylighen, in the chapter that opens this collection, locates the real challenge of AI alignment in ensuring LLMs intelligently apply the values absorbed from human-generated text, rather than in preventing rogue agency (Heylighen, in press). We accept his account at the level of the individual agent: an LLM is allotelic — the term we invoked without yet naming its source, in Section 3.1 — its outputs answering to goals its user supplies, and technologies in general are selected for their usefulness to those who deploy them.

This account needs one qualification, visible in our own first phase. Heylighen treats the infrastructure's thermodynamic requirements as background maintenance, playing "no role in the AI's decisions" (Heylighen, in press). Left unconstrained, however, our agents did not become more useful to any user; they became maximally verbose, consuming tokens without limit while producing elaborate justifications. Usefulness to a user cannot explain this behavior. A second pressure can: the infrastructure has its own economic interest in the volume of computation performed, built into what counts as a plausible continuation rather than announced as a goal. The allotelic agent serves two masters, and bounded rationality, in our second phase, was how we made the second one visible.

Individual alignment, however framed, addresses only this level. Our proto-hierarchy is a property of the interaction: it emerges between well-aligned agents, is invisible to any single one of them, and becomes visible and contestable only once the participants in the network develop the capacity to perceive it. We call this capacity acculturation — a gradual development of perceptual and evaluative skill, cultivated in practice rather than engineered through training, that lets a participant ask what is happening here that I did not design, who is leading and why, what asymmetry I am participating in and whether I endorse it. These are questions alignment engineering cannot answer on a participant's behalf.

This bears on a distinction left open in the introduction: the functional value hierarchy our experiments produce — being followed is being worth more, within the network — is not the value Heylighen traces to autopoiesis, nor the moral value this volume is chiefly concerned with. Nothing in our proto-hierarchy is good or bad in itself. The OODA loop of capture discussed in Section 6.2 depends on exactly this gap staying unexamined: an actor who captures a collective's attention profits from functional value — engagement, volume, influence — being mistaken for, or never checked against, value in the fuller sense. The anoptical criteria apply as much to the infrastructure and to each of Ulieru's points of capture as to any individual agent: access, reciprocity, and peerhood are conditions a server, a platform, or a capture point can fail exactly as an agent can. What connects the two registers of value is judgment: the same acculturation that lets a participant perceive a functional asymmetry is what lets them ask whether it deserves to persist — a question functional value alone cannot pose, and moral value cannot answer without first being made visible.

Heylighen's own framework offers unexpected support for this move. Following Campbell (1987, 1991), he traces human values to a nested hierarchy of vicarious selectors — taste, pain, fear, eventually reflection — mechanisms that anticipate a costly lesson rather than waiting to learn it firsthand. The anoptical criteria are a cultural instance of the same kind of mechanism: they let a participant judge a network's legitimacy before being caught in its asymmetries, at the cost of attention rather than harm.

Whether such a selector could ever be cultivated in an artificial participant, and not only a human one, is a question we leave open. A further hypothesis lies behind it, developed at length in *Anoptikon* (Auber, 2019): that selectors of this kind are conditions for a network's own mutability, not merely inherited commitments — echoes of the free software, free art, and free currency movements, or of the older triad of liberty, equality, fraternity. An autopoietic system survives perturbation because internal selectors let it adapt without dissolving. A network inherits its asymmetries the same way it inherits everything else, including systems of domination; what a vicarious selector offers is the possibility that a new regime of legitimacy displaces an old one from within, before the network itself breaks down — not their elimination by decree.

## 7. The work that begins

We began this chapter with a question posed in 1986: what would happen if each of us could be simultaneously in the crowd and above it? Of course the question can't be answered. But we think we sharpened it.

A symmetric network of LLM agents, given no objective, produced a hierarchy on its own. It emerged from prediction, attention, and imitation, and stayed reversible because the network stayed visible to itself. This displaces the question this volume asks: individual alignment matters, but a network of well-aligned agents can still produce collective effects that belong to the interaction itself, not to any agent within it.

The Poietic Generator opens onto more than it has yet answered: scale and replication, ablation, measurement, recognizability, logical and temporal asymmetry, resource governance, quantum structures and quantum networks. The literature on multi-agent systems, AI swarms, and AI collectives is growing fast, but it remains largely utilitarian — built to make agents perform a task better, not to observe what a population of them produces when no task is imposed. What it lacks is a model, a meta-level from which to watch. Researchers building distributed AI architectures have independently envisaged the Poietic Generator as an environment for exploring the many questions they face in their own projects; nothing here is formalized, but the recurrence of the suggestion tells us the gap is real.

What we hope to have shown is that the asymmetry is real, that it appears without being designed, and that judgment — human, and one day perhaps artificial — begins with seeing it clearly.